\documentclass[12pt]{article}

\usepackage[text={6in,8.5in},centering,top=1in]{geometry}

\usepackage{amsmath,amssymb,booktabs,graphicx,epstopdf,xcolor}

\usepackage{url,hyperref}

\usepackage{lineno}

\usepackage[font={small,sf},textfont=sf,labelfont=bf,
width=\textwidth]{caption}

\usepackage[mono=false]{libertine}
\usepackage{libertinust1math}
\usepackage[T1]{fontenc}

\usepackage{authblk}

\newlength{\graph}
\newlength{\betweengraphs}
\title{The evolution of cooperation in a mobile population on random networks: Network topology matters only for low-degree networks}

\author[1]{Igor V.~Erovenko}
\author[2]{Mark Broom}
\affil[1]{Department of Mathematics and Statistics, University of North Carolina at Greensboro, Greensboro, NC 27402, USA}
\affil[2]{Department of Mathematics, City, University of London, Northampton Square, London EC1V 0HB, UK}

\begin{document}

\maketitle

\begin{abstract}
We consider a finite structured population of mobile individuals that strategically explore a network using a Markov movement model and interact with each other via a public goods game. We extend the model of Erovenko~et al.~(2019) from complete, circle, and star graphs to various random networks to further investigate the effect of network topology on the evolution of cooperation. We discover that the network topology affects the outcomes of the evolutionary process only for networks of small average degree. Once the degree becomes sufficiently high, the outcomes match those for the complete graph. The actual value of the degree when this happens is much smaller than that of the complete graph, and the threshold value depends on other network characteristics.
\end{abstract}

\section{Introduction}





The consideration of population structure is often important when trying to build realistic evolutionary models. This has been done in a number of ways, for example using meta-populations \cite{Levins1969} including the island model structure \cite{Constable2014}. A particularly influential methodology is that of evolutionary graph theory \cite{Lieberman2005}. Here individuals reside on vertices connected by edges and interact with their neighbours through playing games. One significant area of research, which is also as central aspect of the current paper, is the evolution of cooperative behaviour. Using the Prisoner’s Dilemma and related games, various results about the evolution of cooperation have been shown, for example that under weak selection on a regular graph, under the death-birth dynamics with selection on death (DBB) cooperation can evolve if the benefit to cost ratio exceeds the graph degree \cite{Ohtsuki2006}.

One disadvantage of evolutionary graph theory is that the games played are naturally pairwise through the edges (although it is also possible to play multiplayer games amongst specific groups). However, real populations involve multiplayer interactions of various types among groups of variable sizes. To facilitate this, a more general and flexible modelling framework was developed \cite{Broom2012, Bruni2014, Broom2015, Broom2021}. Whilst maintaining most of the elegant features of evolutionary graph theory, this framework allowed for populations to form variable sized groups which could involve correlated movement \cite{Broom2020} and also the history of the process. The evolution of cooperative behaviour has also been considered in such populations, and distinctive results included (again for DBB) the powerful benefit to cooperation of small sub-populations \cite{Pattni2017}. In particular, history-dependence was introduced using Markov movement models in \cite{Pattni2018}. Here individuals moved around the territory interacting with whichever other individuals shared their vertex at any time, before returning to their home vertex for the reproduction phase; the precise process is described in Section~\ref{sec:model} of the current paper. In this work, only an underlying complete graph was considered, so potential structural effects were obscured. 

This model was further and more systematically investigated in \cite{Erovenko2019b}, where three underling graphs were considered, the complete graph, the circle graph and the star graph. These three graphs are particularly interesting, as they form extremes for two key topological properties of graphs, the clustering coefficient and degree centralization. The complete graph, circle graph and star graph have clustering coefficients and degree centralizations of (1, 0), (0, 0) and (0, 1) respectively. They exhibited very different evolutionary outcomes as we outline below, and one conjecture is that the two aforementioned graph properties are central to which type of outcomes result for graphs more generally. In common with \cite{Pattni2017} a multi-player public goods games (the charitable prisoner’s dilemma, see \cite{Broom2019}) was considered where an individual's strategy was comprised of two components, its interactive strategy (defect or cooperate) and its staying propensity (influencing how much it moved around the graph). Movement around the graph incurred a cost, and the size of this cost was an important factor. Two population scenarios were considered; mixed populations, representing situations where mutations occurred in either strategy with equivalent frequency and mutant-resident populations representing the alternative scenario, where mutation in the movement strategy was much more likely.

 In \cite{Erovenko2019b} we observed that for mixed populations cooperators did better on the complete graph and worse on the star graph. The cooperation strategy outperformed the defection one for all but the largest movement costs on the complete graph, for small (but non-zero) to intermediate costs on the circle, and (to a small extent) for intermediate movement costs on the star graph, with otherwise defectors performing better. For mutant-resident populations, the defectors could not replace a cooperator population (i.e., the fixation probability was less than neutral drift) on a complete graph (for large enough populations), could replace cooperators only for sufficiently high movement costs for the circle graph, and could replace cooperators in all cases on the star. Mutant cooperators could replace defector populations for sufficiently small movement costs for all three graphs. Taking all together, we saw that cooperation performed best on the complete graph and worst on the star. This is an interesting difference to the evolutionary graph theory case described previously, where the lower degree graphs were more favourable for cooperation.

 An important message from \cite{Erovenko2019b} was that the stability of a population of defectors is determined by the movement cost, whilst the stability of a population of cooperators is determined by the network topology. In the current paper, we build upon the above work by considering a more general and realistic set of underlying graphs, looking at the effect of network topology beyond the extreme cases. 
 
The above three types are very regular in structure, and so some of the properties exhibited could be influenced by that regularity. What about more randomly generated graphs? Here we consider different classic methods of generating random graphs for the underlying structure upon which our population will move. In particular Barabási--Albert, Erdős--Rényi, random regular and Watts-Strogatz networks. Similarly it was suggested in \cite{Erovenko2019b} that the clustering coefficient and degree centralization could be important properties for the evolution of cooperation. In this paper, we also consider both the average degree and the average shortest path length. More complex graphs of this type had previously been considered in \cite{Schimit2019, Schimit2022} for the independent model, and it had been observed that there could be significant differences between fixation probabilities for different types of graphs, although the different dynamics considered had little effect. In a forthcoming paper \cite{Pires2023}, we show that it is the network topology and movement cost rather than the replacement mechanism that determine the outcome of the evolution of cooperation in a Markov movement model from \cite{Erovenko2019b}. In the current paper we see that the different network types have their own character when they have a low degree, but as the degree of the graphs increase, they typically resemble the results for the complete graph from \cite{Erovenko2019b}. After an introduction of the model in section \ref{sec:model} these results are demonstrated in detail in sections \ref{sec:rare} and \ref{sec:nonrare} and discussed in section \ref{sec:discuss}.

\section{Model}\label{sec:model}

We start with a brief description of a particular instance of the general framework of Broom and Rycht\'{a}\v{r} \cite{Broom2012} for modeling multiplayer interactions in finite structured networks. For further technical details we refer the reader to \cite{Erovenko2019b} on which the current paper builds. Then we explain how the model from \cite{Erovenko2019b} has been adapted to accommodate any network and how we implemented the process on a variety of random networks.

\subsection{The Markov model of the evolution of cooperation on multiplayer networks with costly movement}

We consider a finite population of $N$ individuals that interact over a network with $N$ nodes. Each node in the network is designated as a home place for a unique individual. Individuals move around the network and interact with those they meet in the same location at any time step via a version of the public goods game. At any moment, a group of arbitrary size and composition may potentially form at any node of the network, and hence we use a multiplayer game with an arbitrary number of players to derive individual payoffs. The individuals in our model possess two independent traits: the interactive strategy in the multiplayer game (cooperator or defector) and the exploration strategy. The exploration strategy is determined by the individual's staying propensity, which is the probability the individual is going to stay at the current place if it is alone.

We initialize the environment by placing all individuals at their home locations and resetting their fitness to zero. Then we start an exploration phase where each individual decides to stay at the current location or to move to one of the neighboring locations. The probability that an individual $I_n$ is going to stay at its current location is determined by its staying propensity $\alpha_n$ and the composition of the group $\mathcal G_n$ of the individual $I_n$. This group is defined as the set of all individuals in the population present at the same location as individual $I_n$ at the current time. Each individual evaluates the attractiveness of its current group by adding up the attractiveness of all other members of the group. The attractiveness $\beta_i$ of an individual $I_i$ to others is defined as
\begin{align}
\beta_i =
\begin{cases}
\beta_C & \text{if } I_i \text{ is a cooperator,}\\
\beta_D & \text{if } I_i \text{ is a defector,}
\end{cases}
\end{align}
and we assume that $\beta_C = 1$ and $\beta_D = -1$. In other words, a group with more cooperators than defectors will have a positive attractiveness, and a group with more defectors than cooperators will have a negative attractiveness. We emphasize that the individual does not count its own attractiveness, and hence the attractiveness of the group to an individual which is alone is zero. 

Let $\beta_{\mathcal G_n}$ denote the attractiveness of the group $\mathcal G_n$ of the individual $I_n$ to that individual. Then the probability that this individual is going to stay at its current location is given by
\begin{align}\label{eq:sigmoid}
h_n \left( \mathcal G_n \right) =
\frac{\alpha_n}{\alpha_n + \left( 1-\alpha_n \right) S^{\beta_{\mathcal G_n}}}
\end{align}
where $0 < S < 1$ is the sensitivity to the group composition parameter. Values of $S$ close to $0$ correspond to high sensitivity, that is, an individual is likely to move away from an unattractive group or to stay in an attractive group regardless of its inherent staying propensity $\alpha_n$. Values of $S$ close to $1$ correspond to low sensitivity to the group composition, and the individual's decisions to stay or to move will be mostly determined by its staying propensity. We assume $S = 0.03$, and hence the individuals are highly sensitive to the composition of their current group. Note that when $\beta_{\mathcal G_n} = 0$, in particular, if an individual is alone, then \eqref{eq:sigmoid} simplifies to $h_n \left( \mathcal G_n \right) = \alpha_n$, as expected. Figure~1 in \cite{Erovenko2019b} shows the plots of the staying probability as a function of group attractiveness for several fixed values of the staying propensity of the individual. For example, when the group attractiveness becomes $3$ (or $-3$), then the individual will stay in (or leave) the current group with probability close to $1$ regardless of its staying propensity.

If an individual decides not to stay at the current location, then it's going to move to one of the adjacent locations (nodes in the network). The location to which an individual is going to move is determined randomly with uniform probability. The only strategic decision each individual faces is whether to stay at the current location. The individuals do not evaluate the adjacent locations to see if any of them might provide more beneficial interactions. See \cite{Erovenko2016, Erovenko2019a, Weishaar2022} for a different approach where individual sample all locations to which they may move. All individuals make these decisions independently of each other; see \cite{Broom2020} for some models of coordinated movement in the population. Additionally, movement in our model is costly; each individual pays a fixed cost $\lambda$ every time they move from their current location.

Once all individuals in the population have had an opportunity to stay or move, we play one round of the multiplayer game (public goods game, in our case). The multiplayer game takes place within each group of individuals that are located at the same node. All individuals receive a base payoff of $1$, which ensures that the fitness of individuals remains positive. Each cooperator pays a cost $c = 0.04$ and produces a reward $v = 0.4$ that is shared by all other members of the group (except itself) equally. Defectors do not incur the cost of production of a public good and do not produce a public good, but they do share in the goods produced by cooperators. The payoff to the individual $I_n$ whose current group is $\mathcal G_n$ is thus given by
\begin{align}
R_{n} \left( \mathcal G_n \right) =
\begin{cases}
1 + \frac{|\mathcal G_n|_C - 1}{|\mathcal G_n| - 1} v - c
& \text{if } I_n \text{ is a cooperator and } |\mathcal G_n| > 1,\\
1 - c
& \text{if } I_n \text{ is a cooperator and } |\mathcal G_n| = 1,\\
1 + \frac{|\mathcal G_n|_C}{|\mathcal G_n| - 1} v
& \text{if }  I_n \text{ is a defector and } |\mathcal G_n| > 1,\\
1
& \text{if } I_n \text{ is a defector and } |\mathcal G_n| = 1,
\end{cases}
\end{align}
where $|\mathcal G_n|_C$ is the number of cooperators in the group $\mathcal G_n$. In particular, a lone cooperator has a smaller payoff than a lone defector. The fitness of the individual $I_n$ obtained at this step is then
\begin{align}
f_n 
=
\begin{cases}
R_{n} \left( \mathcal G_n \right) - \lambda & \text{ if the individual moved}\\
R_{n} \left( \mathcal G_n \right) & \text{ if the individual stayed}.
\end{cases}
\end{align}

After one round of the multiplayer game has been played with the current distribution of the population on the network, each individual again decides whether to stay at the current location or to move to one of the random neighboring locations. This is followed by another round of the multiplayer game with possibly different group compositions. This move-play sequence repeats for a fixed number of discrete steps $T$, called the exploration time. At the end of this exploration phase, all individuals instantaneously return to their initial home places, and their fitness is the total fitness accumulated over $T$ rounds of the public goods game. Throughout this paper we shall set $T=10$, in common with the baseline value in the previous work. 

We evolve the population using the birth-death-birth (BDB) dynamics \cite{Masuda2009b}. With these dynamics, an individual for reproduction (birth) is chosen first with probability proportional to the accumulated fitness, and an individual for replacement (death) is then chosen with probability proportional to the \emph{replacement weights}. The replacement weights can be thought of as frequencies of local interactions, and they are defined as follows. Since reproduction occurs locally at the individuals' home places, we consider all possible groups that may form when each individual is given a chance to move once from its home place. Recall that when an individual is alone, then its probability of staying at the current (home) place is equal to its staying propensity. If an individual ends up alone, then it spends a unit of time with itself, and hence it can only replace itself. Otherwise the individual allocates a unit of time equally between all other individuals in its group (not including itself). We then take a weighted sum of these time allocations multiplied by the probabilities of the given groups forming. We were able to compute the replacement weight analytically in \cite{Erovenko2019b} due to the inherent symmetry in the complete, circle, and star graphs. We will explain in the next section how the replacement weights can be handled for arbitrary networks.

In classical evolutionary graph theory, these replacement weights are static, and they are associated with the edges of the graph. In our model, the replacement weights depend on the staying propensities of the individuals comprising the population, and hence they evolve with the population. Therefore, we have two different interconnected structures in our model:
\begin{itemize}\itemsep0em
\item The static network (interaction network) over which individuals move and interact via a multiplayer game

\item The dynamic network (evolutionary graph) whose structure is derived from that of the interaction network, and whose edge weights evolve with the population
\end{itemize}
See \cite{Erovenko2019b} for a more detailed discussion of this decoupling of the interaction and replacement structures in our model as well as in other models \cite{Ohtsuki2007}.

We assume that the natural selection process works on a faster time scale than mutations, and hence at any time there are at most two types of individuals in the population. The type of an individual is determined by its interactive strategy in the public goods game (cooperator or defector) and its staying propensity. For example, we may have two types that are both cooperators but having different staying propensities, or cooperators and defectors which may have either identical or different staying propensities. Mutations may affect the interactive strategy, or staying propensity, or both traits simultaneously. We therefore consider two scenarios with different time scales of mutations of these two traits. They are explained in the corresponding sections later.

After a replacement even takes place, we reset the fitness of all individuals to zero, and start a new exploration and fitness accumulation phase. The process continues until only one type remains in the population.

A comprehensive summary of the model parameters and their values is shown in table~\ref{table:parameters}. The parameter values correspond to the baseline values used in \cite{Erovenko2019b}.

\begin{table}[!h]
\caption{Model parameters.}\label{table:parameters}\centering
\begin{tabular}{clr}
\toprule

\bfseries Notation & \bfseries Meaning & \bfseries Values\\

\midrule

$N$ & Population size & $50$ \\

$T$ & Exploration time & $10$\\

$\lambda$ & Movement cost & $\{ 0, 0.1, 0.2, \dots, 0.8, 0.9 \}$\\

$\gamma$ & Cooperator staying propensity & $\{ 0.01, 0.1, 0.2, \dots, 0.9, 0.99 \}$\\

$\delta$ & Defector staying propensity & $\{ 0.01, 0.1, 0.2, \dots, 0.9, 0.99 \}$\\

$c$ & Cost of cooperation & $0.04$\\

$v$ & Reward of cooperation & $0.40$\\

$S$ & Sensitivity to group members & $0.03$\\

$\beta_C$ & Cooperator attractiveness & $1$\\

$\beta_D$ & Defector attractiveness & $-1$\\

\bottomrule
\end{tabular}
\end{table}

\subsection{Modeling the Markov process on random networks}

We implemented a Monte Carlo simulation of the Markov process described above on four types of random networks with 50 nodes: (1) Barabási--Albert networks; (2) Erdős--Rényi networks; (3) random regular networks; and (4) Watts--Strogatz networks. The size of the networks matches the largest size of the networks considered in \cite{Erovenko2019b}. An exact stochastic simulation of this process is computationally expensive, and running the simulations on an HPC cluster and analyzing data for a project like this one takes approximately 6 months. This is the main limitation to our ability to handle larger networks with the current technologies.

\begin{figure}
\centering
\small (a) \includegraphics[width=0.4\textwidth]{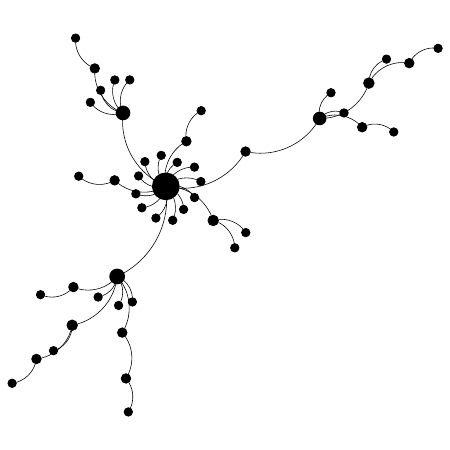} 
\hspace{0.2in} 
(b) \includegraphics[width=0.4\textwidth]{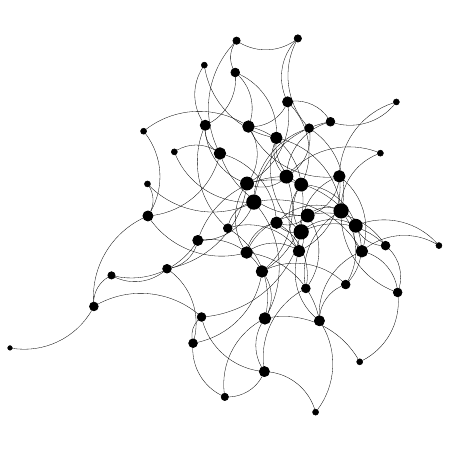}\\[0.1in]
(c) \includegraphics[width=0.4\textwidth]{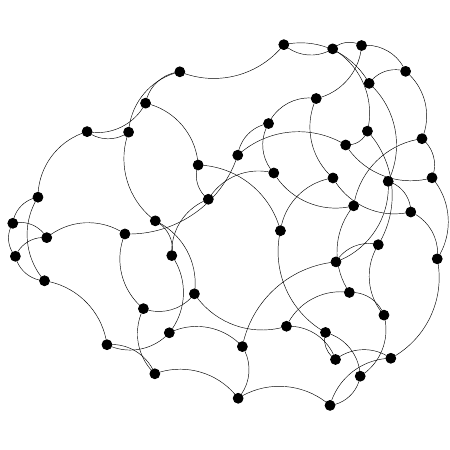} 
\hspace{0.2in}
(d) \includegraphics[width=0.4\textwidth]{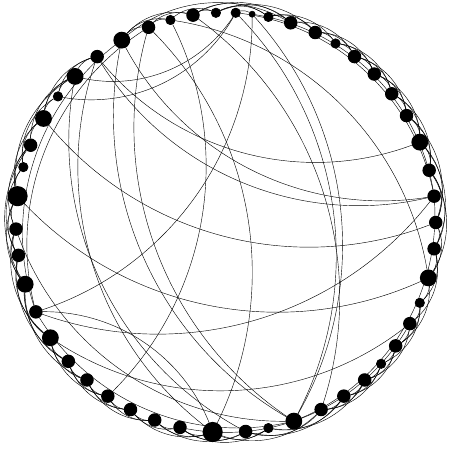}
\caption{Sample random networks with 50 nodes among those we consider. (a) Barabási--Albert network with $d=1$; (b) Erdős--Rényi network with $p=0.1$; (c) random regular network with $d=3$; (d) Watts--Strogatz network with $d=6$ and $p=0.1$.}
\label{fig:networks}
\end{figure}

The networks (1)--(3) are defined by a single parameter, while the network (4) is defined by two parameters. All these parameters fall into two categories: degrees of vertices or probabilities of certain events during the network construction process. We considered 10 values of each network parameter, and generated 10 sample networks for each fixed parameter value using the \texttt{networkx} package. Below is a brief description of the parameters for each network and values that were used to generate sample networks.
\begin{itemize}\itemsep0em
\item A Barabási--Albert network is constructed using a preferential attachment algorithm \cite{Barabasi1999}, and the parameter is the degree $d = 1, 2, \dots, 10$ of each new node.

\item An Erdős--Rényi network is constructed by taking each pair of nodes and joining them with a link with a fixed probability $p = 0.1, 0.2, \dots, 1$. Note that for $p=1$ we obtain a complete graph.

\item A random regular network is constructed to ensure that each node has the same fixed degree $d = 3, 4, \dots, 12$.

\item A Watts--Strogatz network \cite{Watts1998} is constructed by starting with a ring lattice where each node has a fixed even degree $d = 4, 6, 8, \dots, 22$ and then rewiring each link with a fixed probability $p = 0, 0.1, 0.2, \dots, 0.9$.
\end{itemize}
Figure~\ref{fig:networks} demonstrates one sample network of each type.

\begin{figure}
\centering
\includegraphics[width=0.64\textwidth]{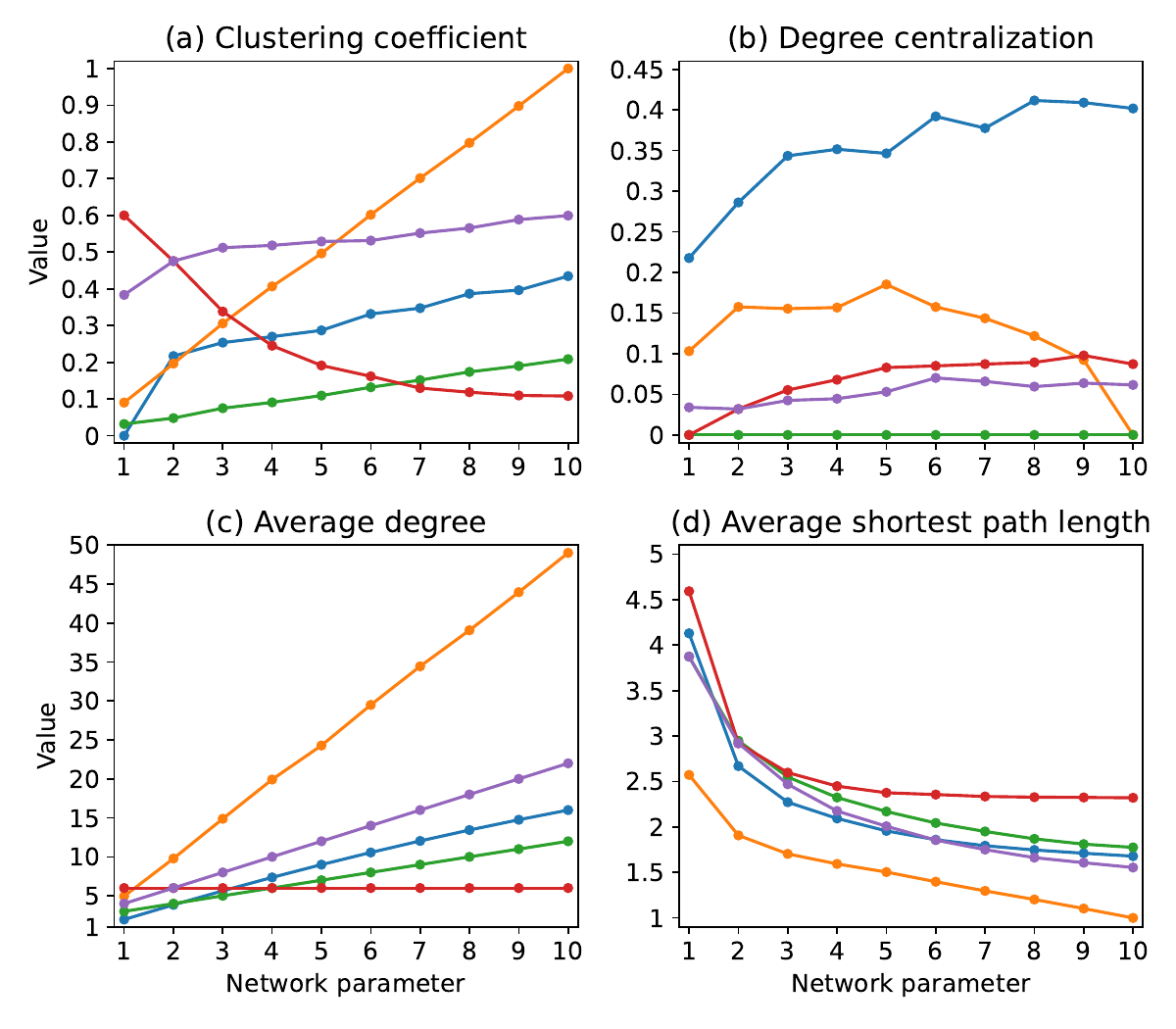} 
\caption{Values of the clustering coefficient, degree centralization, average degree, and average shortest path length for our random networks. Color code: blue---Barabási--Albert network, orange---Erdős--Rényi network, green---random regular network, red---Watts--Strogatz network with fixed $d=6$ and varying $p$, violet---Watts--Strogatz network with fixed $p=0.1$ and varying $d$.}
\label{fig:topchar}
\end{figure}

In \cite{Erovenko2019b} we hypothesized that the clustering coefficient and degree centralization were two of the network topology characteristics that were responsible for the stability of the population of cooperators. In this paper, we also look at the average degree and the average shortest path length; figure~\ref{fig:topchar} shows the values of these four network topology characteristics for all random networks we consider. These values are computed from the actual sample networks we generated rather than the average expected values for the networks of a given type.

The main difference between the implementation of the exact stochastic simulation of the process in the current project and that in \cite{Erovenko2019b} is that in \cite{Erovenko2019b} we used analytically computed replacement weights for each possible population distribution. While there are $2^N$ possible states for the distribution of a population consisting of two types of individuals over a network with $N$ nodes, the complete, circle, and star networks that were handled in \cite{Erovenko2019b} have strong symmetry, which made it possible to reduce the number of possible states down to $N$, $32$, and $2N$ for the complete, circle, and star networks respectively. This cannot be done for random networks, and hence we had to simulate the replacement weights.

The replacement weights were simulated as follows. At the reproduction phase, each individual was given an opportunity to either stay at its home location with probability equal to its staying propensity or to move at a random neighboring location with uniform probability. In other words, the individual $I_n$ stayed home with probability $\alpha_n$ and moved to each of the neighboring locations with probability $\left( 1-\alpha_n  \right) / d$ where $d$ is the degree of the home location node for individual $I_n$. If an individual chosen for reproduction found itself alone, then its offspring replaced the individual itself and there was no change in the population composition and distribution. If an individual chosen for reproduction found itself in a group with at least one other individual, then it replaced one other individual from the group with uniform probability.

\section{Results: rare interactive mutations case}\label{sec:rare}

In this scenario, we assume that the mutation rate of interactive strategies is much slower than the mutation rate of staying propensities. We start with a resident population of cooperators or defectors using the same exploration strategy (i.e., having the same staying propensity). We then introduce a mutant that differs from the resident population in staying propensity but not interactive strategy. The resident population thus evolves to the optimal (Nash equilibrium) staying propensity given the network structure and movement cost. Defectors have no incentive to move in the absence of cooperators, and hence the resident defectors always have $0.99$ as the optimal staying propensity. 

By the time an interactive strategy mutant appears in the population, the residents will have evolved to the optimal staying propensity. The interactive strategy mutant may also have a different staying propensity, and hence we simulate the invasion of the resident population by mutants having all possible staying propensities. We run a Monte Carlo simulation of the natural selection process to estimate the mutant fixation probability $\rho$. We compare the estimated mutant fixation probability with the neutral drift fixation probability $\rho = 1/N = 0.02$. We will denote the fixation probability of the fittest mutant cooperator by $\rho^C$ and the fittest mutant defector by $\rho^D$. Using the terminology from \cite{Taylor2004}, there are four possible outcomes of the selection process:
\begin{itemize}\itemsep0em
\item Selection favors cooperators when $\rho^C > 1/N$ and $\rho^D < 1/N$

\item Selection favors defectors when $\rho^C < 1/N$ and $\rho^D > 1/N$

\item Selection favors change when $\rho^C > 1/N$ and $\rho^D > 1/N$

\item Selection opposes change when $\rho^C < 1/N$ and $\rho^D < 1/N$
\end{itemize}

To estimate the mutant fixation probabilities we average the results of $n = 100,000$ independent trials for each combination of parameters. These trials are split into ten groups of 10,000---one group for each sample random network of the given type. Using the binomial distribution, the standard deviation of the simulated fixation probability is given by $\sqrt{\rho \left( 1-\rho \right) / n}$, where $\rho$ is the actual fixation probability. The stochastic error is critical when the fixation probability is close to the neutral one. Assuming $\rho = 1/N = 0.02$, the standard deviation is equal to $0.00044$. We adopt the following convention:
\begin{itemize}\itemsep0em
\item We assume that selection favors the mutant if the mutant fixation probability exceeds the neutral one by at least two standard deviations
\end{itemize}
With this convention, the simulated mutant fixation probability is considered to exceed the neutral one if it is greater than $0.02088$.

We now proceed to analyzing the outcomes in this scenario for different networks. We draw parameter regions that correspond to four possible outcomes and plot fixation probabilities of the fittest cooperator and defector mutants. As the parameters change in discrete steps, whenever an outcome changes we draw the dividing line halfway between the parameter values. The color codes for the regions are as follows:
\begin{itemize}\itemsep0em
\item Blue: selection favors cooperators

\item Orange: selection favors defectors

\item Yellow: selection favors change

\item Gray: selection opposes change
\end{itemize}
In the plots of mutant fixation probabilities, the thick gray line indicates the area of stochastic uncertainty around the neutral fixation probability. It is centered at the neutral fixation probability, and its thickness is equal to four standard deviations. We also apply the same scale for the mutant fixation probabilities plots across different networks for convenience of comparison.

One common theme that we observe for all networks is that once the average degree becomes sufficiently high, the qualitative outcomes are identical to those for the complete graph of size $N=50$ from \cite{Erovenko2019b}. We recall that the complete graph is characterized by the highest clustering coefficient ($1$), lowest degree centralization ($0$), highest average degree ($N-1$), and lowest average shortest path length ($1$). 




For networks with small average degrees, the outcomes resemble those for either circle or star graphs from \cite{Erovenko2019b}. For reader's convenience we provide the ranges of movement costs which correspond to one of the possible outcomes for each of these three basic graphs in Table~\ref{table:basicrareresults}. Applying our conventions for drawing parameter regions for the complete graph as an example, we have a blue region extending from the left border (movement cost $0$) up to movement cost $0.25$ (halfway between the parameter values where the qualitative change takes place) and a gray region extending from movement cost $0.25$ up to the right border (movement cost $0.9$).

\begin{table}[!h]
\caption{Ranges of movement costs corresponding to each possible selection outcome in the rare interactive mutations scenario for complete, circle, and star graphs of size $50$ from \cite{Erovenko2019b}.}\label{table:basicrareresults}\centering
\begin{tabular}{lcccc}
\toprule

\bfseries Network & \bfseries Favors Coop. & \bfseries Favors Def. & \bfseries Favors change & \bfseries Opposes change\\

\midrule

Complete & $0$--$0.2$ & $-$ & $-$ & $0.3$--$0.9$\\
Circle & $0$--$0.3$ & $0.4$--$0.9$ & $-$ & $-$\\
Star & $-$ & $0.4$--$0.9$ & $0$--$0.3$ & $-$\\

\bottomrule
\end{tabular}
\end{table}

\subsection{Barabási--Albert networks}

\begin{figure}
\centering
\includegraphics[width=\textwidth]{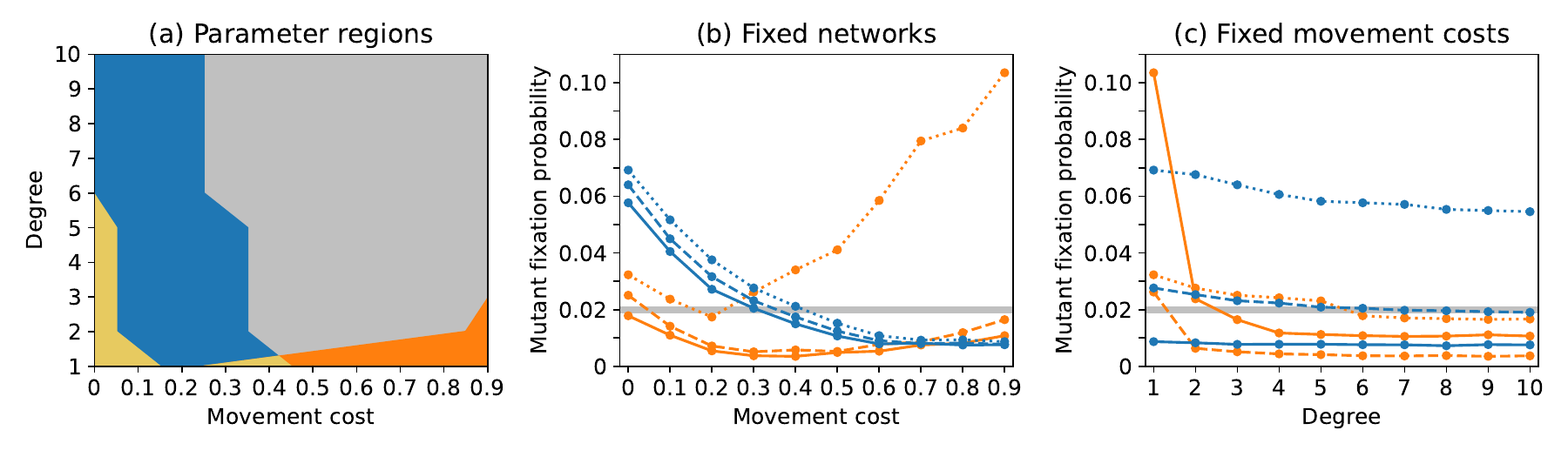} 
\caption{The outcomes in the rare interactive mutations scenario for Barabási--Albert networks. (a) Parameter regions showing the possible outcomes. The results match those of the complete graph starting from $d=6$.  (b) Fittest cooperator (blue) and defector (orange) mutant fixation probabilities as functions of the movement cost. Each plot corresponds to a fixed value of the network parameter: $d=1$ for dotted lines, $d=3$ for dashed lines, $d=6$ for solid lines. (c) Fittest cooperator (blue) and defector (orange) mutant fixation probabilities as functions of the network parameter. Each plot corresponds to a fixed value of the movement cost: $\lambda = 0$ for dotted lines, $\lambda = 0.3$ for dashed lines, $\lambda = 0.9$ for solid lines.}
\label{fig:BAslow}
\end{figure}

Figure~\ref{fig:BAslow} shows the outcomes for Barabási--Albert networks. For $d=1$, the network resembles the hub-and-spoke topology of the star graph because it has few nodes of high degree and most nodes have small degree; see figure~\ref{fig:networks}(a). Consequently, the outcomes in this case are similar to the ones for the star graph \cite{Erovenko2019b}. The only difference is that for the Barabási--Albert network selection favors cooperators for $\lambda = 0.2$, while selection favors change for the star graph. As the degree of newly attached nodes increases, the behavior gradually drifts towards that of the complete graph with the region corresponding to selection opposing change expanding, the regions corresponding to selection favoring defectors and change disappearing, and the region corresponding to selection favoring defectors shrinking until stabilizing at the size identical to that in the complete graph. The qualitative outcomes match those of the complete graph \cite{Erovenko2019b} starting from $d=6$, which corresponds to the values of clustering coefficient $0.33$, degree centralization $0.39$, average degree $10.56$, and average shortest path length $1.86$ (actual values from our 10 sample networks rather than expected values).

Next, we will compare the actual mutant fixation probabilities in figure~\ref{fig:BAslow}(b) to those on the complete and star graphs. For $d=1$, the fittest mutant cooperator fixation probabilities on the Barabási--Albert network are slightly lower than those on the star graph for movement costs $0$ and $0.1$, but they are approximately the same for all larger movement costs  (see figure~6(c) in \cite{Erovenko2019b}). Yet the fittest mutant defector fixation probabilities differ significantly. On the Barabási--Albert network, these fixation probabilities decrease from $\lambda = 0$ to $\lambda = 0.2$, with the fixation probability at $\lambda = 0.2$ dropping below the neutral threshold, and then increase sharply with the increasing values of the movement cost. But on the star graph, the mutant defector fixation probabilities increase slowly and gradually from $0.04$ for $\lambda = 0$ to $0.05$ for $\lambda = 0.9$.

For $d=6$, the fittest mutant cooperator fixation probabilities on the Barabási--Albert network are slightly higher than those on the complete graph for movement costs up to $0.4$, but they are approximately the same for all larger movement costs  (see figure~2(c) in \cite{Erovenko2019b}). Similarly for the fittest mutant defector fixation probabilities.

Looking at the mutant fixation probability plots in figure~\ref{fig:BAslow}(c), we observe that the mutant cooperator fixation probabilities decrease slightly with the increasing degree of the nodes in the network for small and intermediate movement costs (dotted and dashed lines) and remain independent of the network parameter for large movement costs (solid line). The mutant defector fixation probabilities show a sharp initial decrease for intermediate and large movement costs, and decrease gradually for small movement costs.

\subsection{Erdős--Rényi networks}

\begin{figure}
\centering
\includegraphics[width=\textwidth]{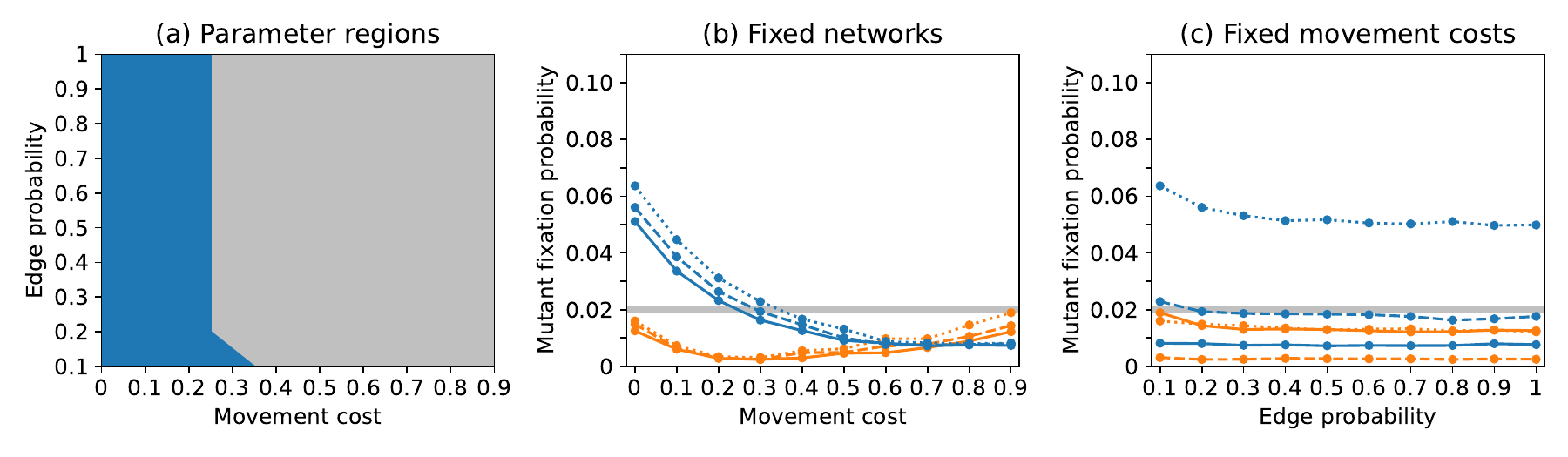} 
\caption{The outcomes in the rare interactive mutations scenario for Erdős--Rényi networks. (a) Parameter regions showing the possible outcomes. The results match those of the complete graph starting from $p=0.2$.  (b) Fittest cooperator (blue) and defector (orange) mutant fixation probabilities as functions of the movement cost. Each plot corresponds to a fixed value of the network parameter: $p=0.1$ for dotted lines, $p=0.2$ for dashed lines, $p=0.8$ for solid lines. (c) Fittest cooperator (blue) and defector (orange) mutant fixation probabilities as functions of the network parameter. Each plot corresponds to a fixed value of the movement cost: $\lambda = 0$ for dotted lines, $\lambda = 0.3$ for dashed lines, $\lambda = 0.9$ for solid lines.}
\label{fig:ERslow}
\end{figure}

Figure~\ref{fig:ERslow} shows the outcomes for Erdős--Rényi networks. The parameter regions exhibit little change with the edge probability parameter, and the qualitative outcomes are identical to those for the complete graph starting from $p=0.2$, which corresponds to the values of clustering coefficient $0.2$, degree centralization $0.16$, average degree $9.79$, and average shortest path length $1.91$. We performed additional computations for $p = 0.11, 0.12, \dots, 0.19$ and discovered that the outcomes match those of the complete graph starting only from $p=0.2$.

Comparing the fittest mutant fixation probabilities in figure~\ref{fig:ERslow}(b) with those for the complete graph (figure~2(c) in \cite{Erovenko2019b}), we observe that the mutant cooperator fixation probability decreases slightly with the increasing edge probability $p$ for all but larger movement costs. The mutant defector fixation probabilities also decrease slightly with the increasing edge probability for small and large movement costs. The fittest mutant fixation probabilities exhibit little change with the network parameter in figure~\ref{fig:ERslow}(c). Most of the change, if any, occurs between $p=0.1$ and $p=0.2$.

\subsection{Random regular networks}

\begin{figure}
\centering
\includegraphics[width=\textwidth]{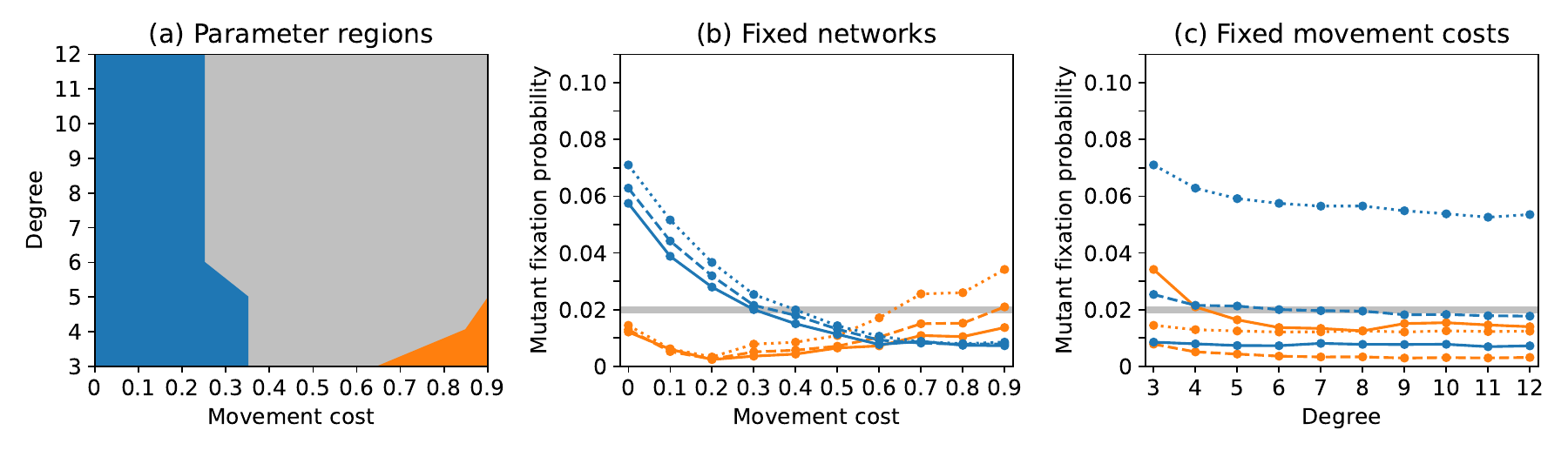} 
\caption{The outcomes in the rare interactive mutations scenario for random regular networks. (a) Parameter regions showing the possible outcomes. The results match those of the complete graph starting from $d=6$.  (b) Fittest cooperator (blue) and defector (orange) mutant fixation probabilities as functions of the movement cost. Each plot corresponds to a fixed value of the network parameter: $d=3$ for dotted lines, $d=4$ for dashed lines, $d=6$ for solid lines. (c) Fittest cooperator (blue) and defector (orange) mutant fixation probabilities as functions of the network parameter. Each plot corresponds to a fixed value of the movement cost: $\lambda = 0$ for dotted lines, $\lambda = 0.3$ for dashed lines, $\lambda = 0.9$ for solid lines.}
\label{fig:RRslow}
\end{figure}

Figure~\ref{fig:RRslow} shows the outcomes for random regular networks. For $d=3$, the outcomes are closest to the ones for the circle graph (a regular network with small average degree) from \cite{Erovenko2019b} with selection favoring cooperators for smaller movement costs and selection favoring defectors for larger movement costs. The only difference is that for the random regular network selection opposes change for intermediate values of the movement cost. As the degree of the nodes increases, the behavior gradually drifts towards that of the complete graph with the region corresponding to selection opposing change expanding, the region corresponding to selection favoring defectors disappearing, and the region corresponding to selection favoring defectors shrinking until stabilizing at the size identical to that in the complete graph. The qualitative outcomes match those of the complete graph starting from $d=6$, which corresponds to the values of clustering coefficient $0.09$, degree centralization $0$, average degree $6$, and average shortest path length $2.32$.

Comparing the fittest mutant fixation probabilities on random regular networks for $d=3$ in figure~\ref{fig:RRslow}(b) with those on the circle graph (figure~4(c) in \cite{Erovenko2019b}), we observe that mutant cooperator fixation probabilities are a bit lower on random regular networks for movement costs up to $0.6$, and mutant defector fixation probabilities are much lower on random regular networks for movement costs exceeding $0.3$.
Comparing the mutant fixation probabilities on random regular networks for $d=6$ with those on the complete graph (figure~2(c) in \cite{Erovenko2019b}), we see that mutant cooperators have a slightly higher fixation probability on random regular networks for movement costs up to $0.5$ while mutant defectors have similar fixation probabilities for all movement costs.
We also notice from figure~\ref{fig:RRslow}(c) that the mutant fixation probabilities either decrease with the increasing degree of the nodes in the network or remain stable.

\subsection{Watts--Strogatz networks}

\begin{figure}
\centering
\includegraphics[width=\textwidth]{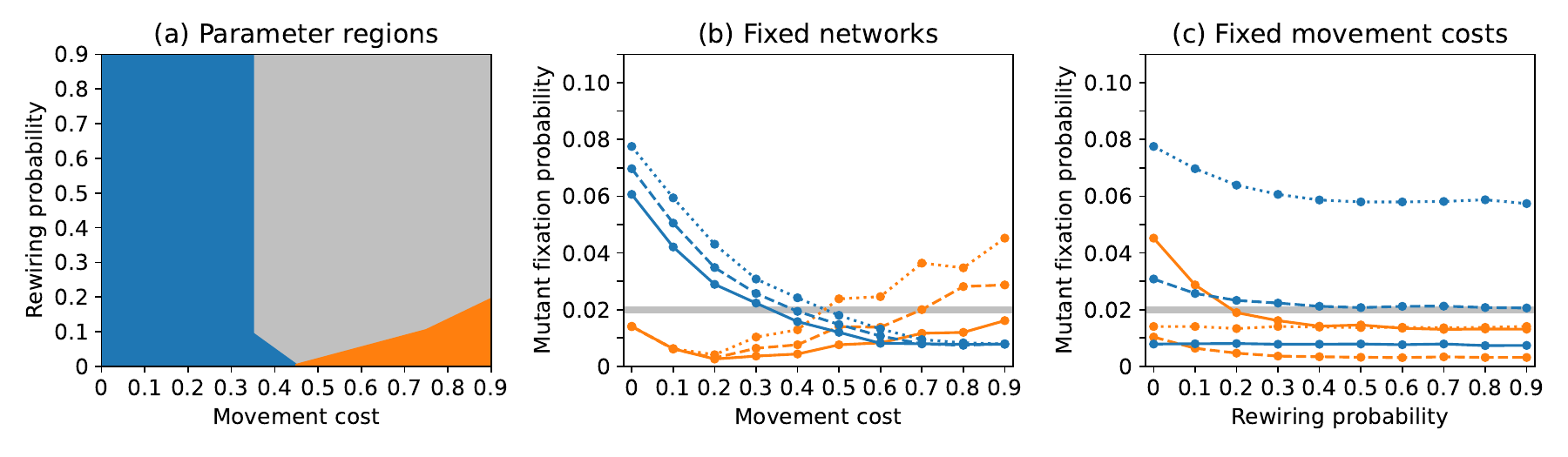} 
\caption{The outcomes in the rare interactive mutations scenario for Watts--Strogatz networks with fixed average degree $d=6$. (a) Parameter regions showing the possible outcomes. The results stabilize starting from $p=0.2$ but never match those of the complete graph. (b) Fittest cooperator (blue) and defector (orange) mutant fixation probabilities as functions of the movement cost. Each plot corresponds to a fixed value of the network parameter: $p=0$ for dotted lines, $p=0.1$ for dashed lines, $p=0.3$ for solid lines. (c) Fittest cooperator (blue) and defector (orange) mutant fixation probabilities as functions of the network parameter. Each plot corresponds to a fixed value of the movement cost: $\lambda = 0$ for dotted lines, $\lambda = 0.3$ for dashed lines, $\lambda = 0.9$ for solid lines.}
\label{fig:WSd6slow}
\end{figure}

Figure~\ref{fig:WSd6slow} shows the outcomes for Watts--Strogatz networks with fixed initial degree $d=6$ of the nodes in the starting ring lattice and varying rewiring probability $p$. When no links in the ring lattice are rewired ($p=0$), the qualitative outcomes are identical to those on the circle graph \cite{Erovenko2019b}: selection favors cooperators for movements costs $0$--$0.4$ and selection favors defectors for movement costs $0.5$--$0.9$. As the rewiring probability is increasing, the region where selection favors defectors disappears and the region where selection favors cooperators shrinks slightly. Starting from $p=0.2$ the regions stabilize, but the outcomes never quite match those of the complete graph because the average degree of the nodes stays constant. Notice in figure~\ref{fig:topchar}(d) how the average shortest path length (red plot) stops decreasing and stabilizes starting from intermediate values of $p$. This could explain why the outcomes never become identical to those for the complete graph in these networks.

Comparing the mutant fixation probabilities for $p=0$ on Watts--Strogatz networks and the circle graph (figure~4(c) in \cite{Erovenko2019b}), we observe that the mutant cooperator fixation probabilities are slightly lower on the Watts--Strogatz network for movement costs up to $0.3$, and the mutant defector fixation probabilities are lower on the Watts--Strogatz network for movements costs starting from $0.4$.
Comparing the mutant fixation probabilities on Watts--Strogatz networks for $p=0.3$ with those on the complete graph (figure~2(c) in \cite{Erovenko2019b}), we see that the mutant cooperator fixation probabilities are higher on the Watts--Strogatz network for movement costs up to $0.5$, and the mutant defector fixation probabilities are slightly higher on the Watts--Strogatz network for small and large movement costs.
We also notice from figure~\ref{fig:WSd6slow}(c) that the mutant fixation probabilities either decrease with the increasing rewiring probability or remain stable.

\begin{figure}
\centering
\includegraphics[width=\textwidth]{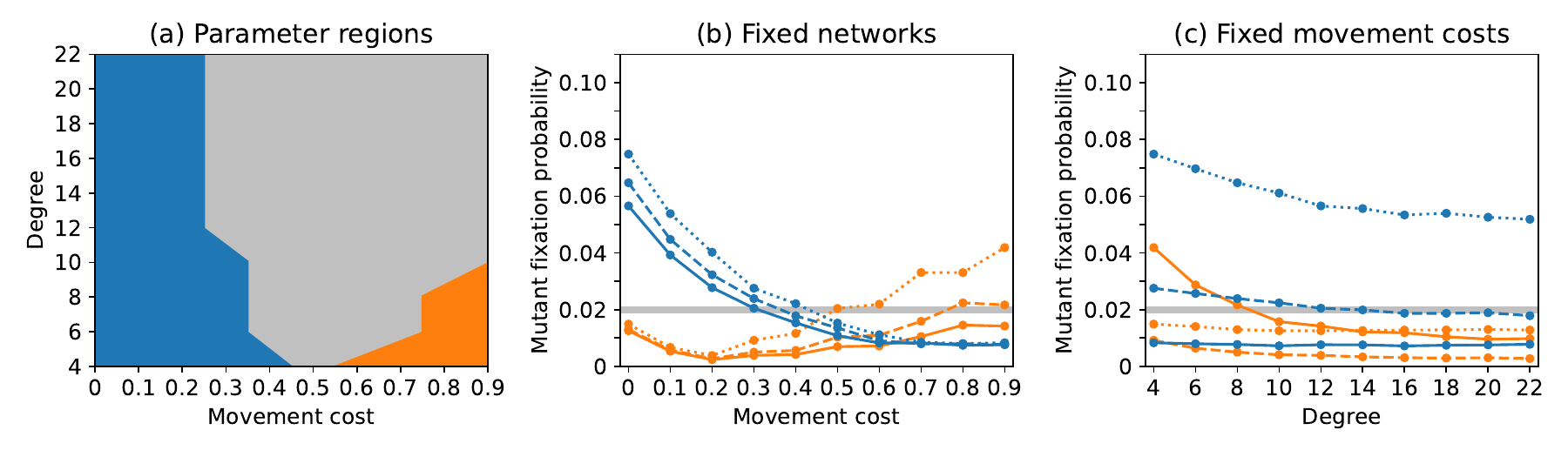} 
\caption{The outcomes in the rare interactive mutations scenario for Watts--Strogatz networks with fixed rewiring probability $p=0.1$. (a) Parameter regions showing the possible outcomes. The results match those of the complete graph starting from $d=12$.  (b) Fittest cooperator (blue) and defector (orange) mutant fixation probabilities as functions of the movement cost. Each plot corresponds to a fixed value of the network parameter: $d=4$ for dotted lines, $d=8$ for dashed lines, $d=12$ for solid lines. (c) Fittest cooperator (blue) and defector (orange) mutant fixation probabilities as functions of the network parameter. Each plot corresponds to a fixed value of the movement cost: $\lambda = 0$ for dotted lines, $\lambda = 0.3$ for dashed lines, $\lambda = 0.9$ for solid lines.}
\label{fig:WSp10slow}
\end{figure}

Figure~\ref{fig:WSp10slow} shows the outcomes for Watts--Strogatz networks with fixed rewiring probability $p=0.1$ and varying initial degree of nodes $d$ in the starting ring lattice. For $d=4$, the outcomes are very close to those on the circle graph. As $d$ is increasing, the regions where selection favors either cooperators or defectors start to shrink while the region where selection opposes change expands. The regions stabilize at $d=12$ and the outcomes match those for the complete graph. This corresponds to the values of clustering coefficient $0.53$, degree centralization $0.05$, average degree $12$, and average shortest path length $2$.

Comparing the mutant fixation probabilities for $d=4$ on Watts--Strogatz networks and circle graph (figure~4(c) in \cite{Erovenko2019b}), we observe that the mutant cooperator fixation probabilities are slightly lower on the Watts--Strogatz network for movement costs up to $0.4$, and the mutant defector fixation probabilities are lower on the Watts--Strogatz network for movements costs starting from $0.4$. This is very similar to what is happening in the $d=6$ case in figure~\ref{fig:WSp10slow}(b).
Comparing the mutant fixation probabilities on Watts--Strogatz networks for $d=12$ with those on the complete graph (figure~2(c) in \cite{Erovenko2019b}), we see that the mutant cooperator fixation probabilities are slightly higher on the Watts--Strogatz network for movement costs up to $0.4$, and the mutant defector fixation probabilities are slightly higher on the Watts--Strogatz network for large movement costs.
We also notice from figure~\ref{fig:WSp10slow}(c) that the mutant fixation probabilities either decrease with the increasing average degree or remain stable.

\subsection{Summary}

Regardless of the network structure, as long as the average degree of the nodes becomes sufficiently high, the qualitative outcomes match those of the complete graph. This is expected, but what was not clear \emph{a priori} is that the actual average degree where this phenomenon is observed is much smaller than that of the complete graph. Moreover, different network topologies require different average degree thresholds starting from which the outcomes become identical. Table~\ref{table:rare-summary} summarizes the values of the network topology characteristics that correspond to the lowest value of the average degree threshold for each network type.

\begin{table}[!h]
\caption{Network topology thresholds for identical outcomes with the complete graph of size $50$.}\label{table:rare-summary}\centering
\begin{tabular}{lrrrr}
\toprule

\bfseries Network & \bfseries Clust.~coeff. & \bfseries Deg.~centr. & \bfseries Avg. deg. & \bfseries Avg. sh. path length\\

\midrule

Complete & $1.00$ & $0.00$ & $49.00$ & $1.00$ \\
Barabási--Albert & $0.33$ & $0.39$ & $10.56$ & $1.86$ \\
Erdős--Rényi & $0.20$ & $0.16$ & $9.79$ & $1.91$ \\
Random regular & $0.09$ & $0.00$ & $6.00$ & $2.32$ \\
Watts--Strogatz & $0.53$ & $0.05$ & $12.00$ & $2.00$ \\

\bottomrule
\end{tabular}
\end{table}

One common feature that these networks share is small average shortest path length. Of course, this alone is not sufficient for having identical qualitative outcomes as the star graph demonstrates \cite{Erovenko2019b}. 

All networks except the Barabási--Albert ones have low degree centralization, which means these networks do not have nodes of large degree surrounded by many nodes of small degree. These hubs are the most likely places for groups of individuals to form. Having few such places means defectors may find it easier to locate and exploit clusters of cooperators, and hence a resident cooperator population is vulnerable to invasion by defectors. The high degree centralization for Barabási--Albert networks is compensated for by a sufficiently high clustering coefficient and average degree. This means cooperators leaving the groups that are being heavily exploited by defectors have a higher chance of forming new clusters at other locations quickly thereafter, while defectors may not be able to find these clusters for some time.

Random regular networks have low values of the clustering coefficient and degree centralization, similarly to the circle graph. Yet they have a higher average degree and much lower average shortest path length than the circle graph. This allows cooperators to ``spread out'' when escaping unattractive groups as well as form new groups. This is much harder to achieve on the circle graph.

\section{Results: non-rare interactive mutations case}\label{sec:nonrare}

In this scenario, we assume that the mutation rate of an individual's interactive strategies is similar to the mutation rate of their staying propensity. Adopting the modeling approach from \cite{Erovenko2019b} for this scenario, we consider a mixed population where half of the individuals are cooperators and half are defectors. We then find the Nash equilibrium staying propensity of each type in this mixed population and record the probabilities that the mixed population evolves to all cooperators or all defectors; these probabilities add up to $1$. We call these probabilities ``fixation probabilities'', but not ``mutant fixation probabilities''. Denoting by $\rho_C$ and $\rho_D$ the fixation probabilities of cooperators and defectors, respectively, we may end up with one of the three possible outcomes (with color codes for parameter regions): 
\begin{enumerate}\itemsep0em
\item Selection favors cooperators if $\rho_C > 1/2 > \rho_D$ (blue color)

\item Selection favors defectors if $\rho_D > 1/2 > \rho_C$ (orange color)

\item Selection is neutral if $\rho_C \approx \rho_D$ (gray color)
\end{enumerate}
To estimate the fixation probabilities, we ran $10,000$ independent trials for each combination of parameters; ten groups of $1,000$ trials for each sample random network. Using the same approach as in the rare interactive mutations case, the standard deviation around the expected fixation probability $1/2$ is equal to $0.005$. Counting two standard deviations in each direction, we assume that selection is neutral when the fixation probability of cooperators $\rho_C$ falls between $0.49$ and $0.51$. The plots of fixation probabilities contain a thick gray line indicating this area of stochastic uncertainty around the neutral fixation probability.

Table~\ref{table:basicnonrareresults} summarizes the outcomes in the non-rare interactive mutations scenario for complete, circle, and star graphs from~\cite{Erovenko2019b}. Similarly to the rare interactive mutations scenario, the outcomes for random networks match those of the complete graph as long as the network has a sufficiently high average degree. The only exception are the Barabási--Albert networks, where selection is neutral for movement cost $0.9$. However, this qualitative difference results from a small quantitative difference in the fixation probabilities. For all networks, the actual fixation probabilities are very close to those on the complete graph for sufficiently large average degrees. In general, increasing the average degree of the network resulted in stable or higher cooperator fixation probabilities.

\begin{table}[!h]
\caption{Ranges of movement costs corresponding to each possible selection outcome in the non-rare interactive mutations scenario for complete, circle, and star graphs of size $50$ from \cite{Erovenko2019b}.}\label{table:basicnonrareresults}\centering
\begin{tabular}{lccc}
\toprule

\bfseries Network & \bfseries Favors Cooperators & \bfseries Favors Defectors & \bfseries Neutral \\

\midrule

Complete & $0$--$0.8$ & $0.9$ & $-$\\
Circle & $0.1$--$0.4$ & $0.5$--$0.9$ & $0$\\
Star & $-$ & $0$--$0.3$ and $0.8$--$0.9$ & $0.4$--$0.7$\\

\bottomrule
\end{tabular}
\end{table}

\subsection{Barabási--Albert networks}

\begin{figure}
\centering
\includegraphics[width=\textwidth]{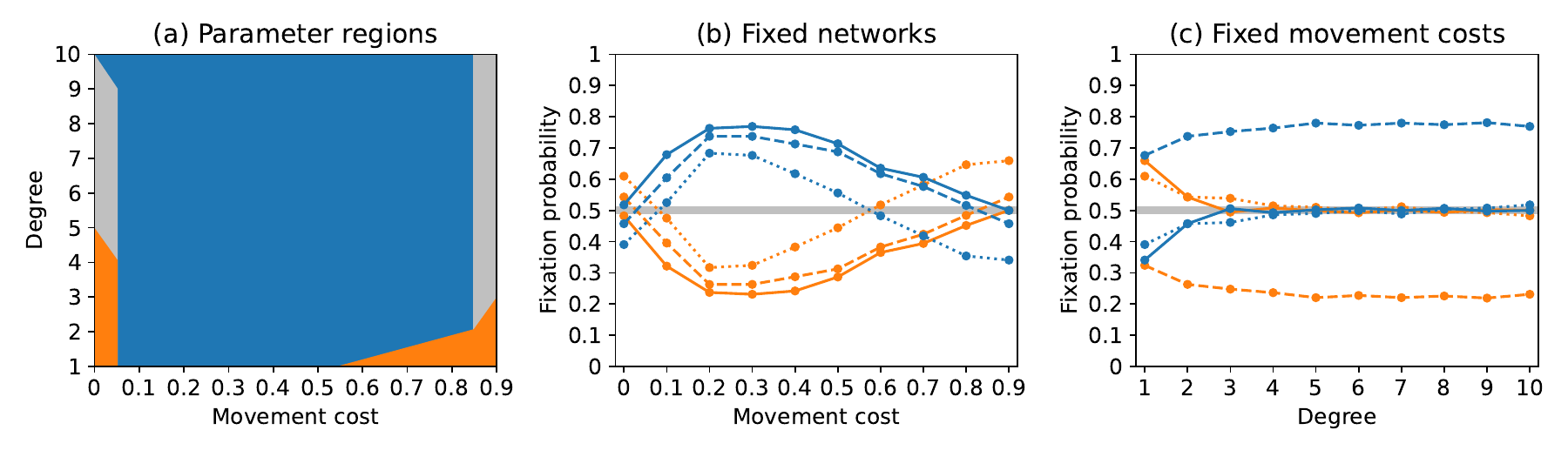} 
\caption{The outcomes in the non-rare interactive mutations scenario for Barabási--Albert networks. (a) Parameter regions showing the possible outcomes. The results almost match those of the complete graph for $d=10$.  (b) Equilibrium fixation probabilities of cooperators (blue) and defectors (orange) as functions of the movement cost. Each plot corresponds to a fixed value of the network parameter: $d=1$ for dotted lines, $d=2$ for dashed lines, $d=10$ for solid lines. (c) Equilibrium fixation probabilities of cooperators (blue) and defectors (orange) as functions of the network parameter. Each plot corresponds to a fixed value of the movement cost: $\lambda = 0$ for dotted lines, $\lambda = 0.3$ for dashed lines, $\lambda = 0.9$ for solid lines.}
\label{fig:BAfast}
\end{figure}

Figure~\ref{fig:BAfast} shows the outcomes for Barabási--Albert networks. For small average degrees, selection favors defectors for small and large movement costs, and selection favors cooperators for intermediate movement costs. As the degree increases, the area where selection favors cooperators expands, while the area where selection favors defectors shrinks and disappears; see figure~\ref{fig:BAfast}(a). For $d=10$, the outcomes have one slight difference from those for the complete graph: for movement cost $0.9$, selection is neutral. We tested what happens for even larger values of $d$, and discovered that this behavior remains stable (up to $d=20$). Even though there is a qualitative difference with the complete graph, this difference stems from a small difference in the actual fixation probabilities because in both cases they are close to the neutral one.

Figure~\ref{fig:BAfast}(b) shows that cooperator fixation probabilities increase with the increasing degree of the networks. Similarly to the complete graph, they are highest for intermediate movement costs independently of the degree (figure~\ref{fig:BAfast}(c)).

\subsection{Erdős--Rényi networks}

\begin{figure}
\centering
\includegraphics[width=\textwidth]{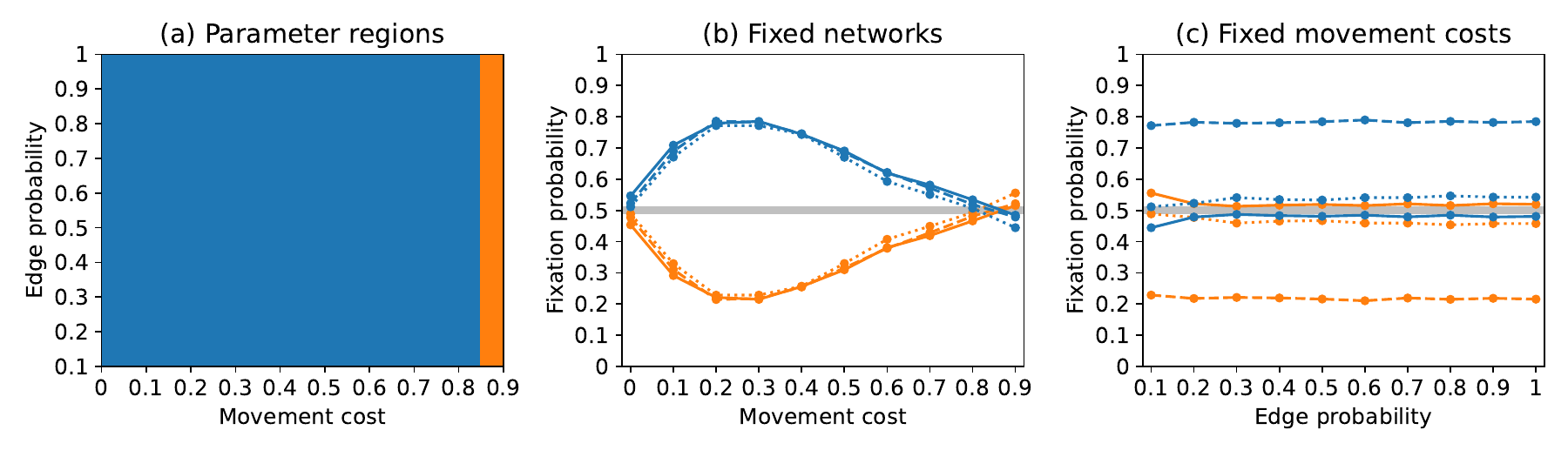} 
\caption{The outcomes in the non-rare interactive mutations scenario for Erdős--Rényi networks. (a) Parameter regions showing the possible outcomes. The results match those of the complete graph starting starting from $p=0.1$.  (b) Equilibrium fixation probabilities of cooperators (blue) and defectors (orange) as functions of the movement cost. Each plot corresponds to a fixed value of the network parameter: $p=0.1$ for dotted lines, $p=0.2$ for dashed lines, $p=0.8$ for solid lines. (c) Equilibrium fixation probabilities of cooperators (blue) and defectors (orange) as functions of the network parameter. Each plot corresponds to a fixed value of the movement cost: $\lambda = 0$ for dotted lines, $\lambda = 0.3$ for dashed lines, $\lambda = 0.9$ for solid lines.}
\label{fig:ERfast}
\end{figure}

Figure~\ref{fig:ERfast} shows the outcomes for Erdős--Rényi networks. These outcomes are qualitatively identical to those for the complete graph throughout the entire range of the network parameter value (figure~\ref{fig:ERfast}(a)). The actual fixation probabilities also show little change with the network parameter (figure~\ref{fig:ERfast}(b)). In this case, the fixation probabilities for $\lambda = 0.9$ were very close to the neutral one, and we ran additional simulations to decrease the size of the stochastic error. With smaller standard deviation, we were able to conclude that selection favors defectors in all cases.

\subsection{Random regular networks}

\begin{figure}
\centering
\includegraphics[width=\textwidth]{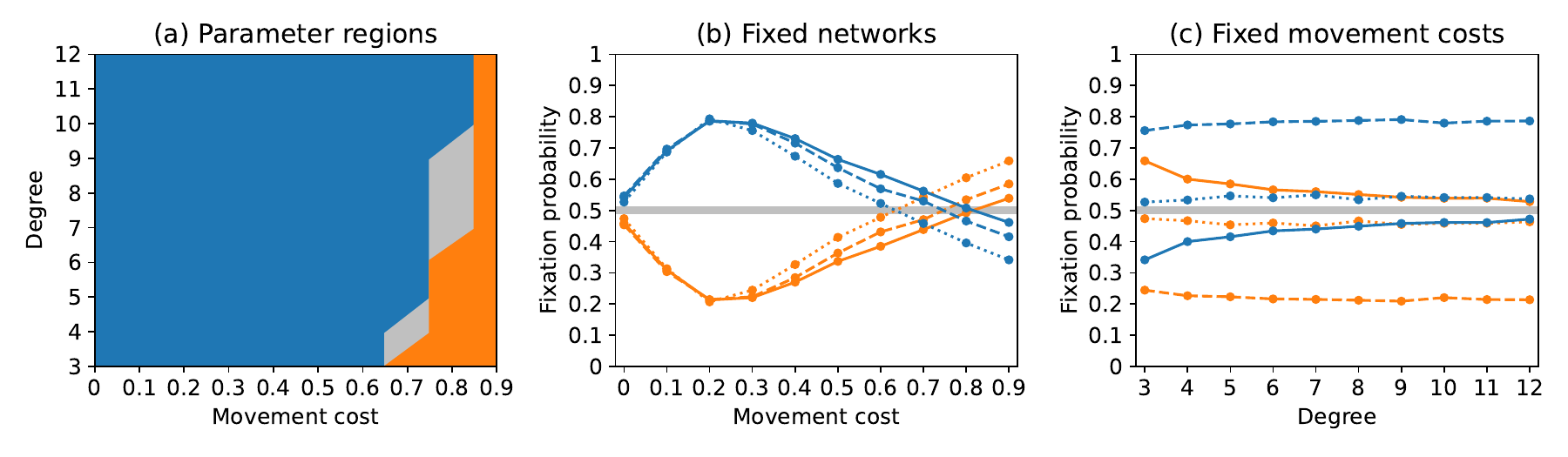} 
\caption{The outcomes in the non-rare interactive mutations scenario for random regular networks. (a) Parameter regions showing the possible outcomes. The results match those of the complete graph starting from $d=10$.  (b) Equilibrium fixation probabilities of cooperators (blue) and defectors (orange) as functions of the movement cost. Each plot corresponds to a fixed value of the network parameter: $d=3$ for dotted lines, $d=5$ for dashed lines, $d=10$ for solid lines. (c) Equilibrium fixation probabilities of cooperators (blue) and defectors (orange) as functions of the network parameter. Each plot corresponds to a fixed value of the movement cost: $\lambda = 0$ for dotted lines, $\lambda = 0.3$ for dashed lines, $\lambda = 0.9$ for solid lines.}
\label{fig:RRfast}
\end{figure}

Figure~\ref{fig:RRfast} shows the outcomes for random regular networks. For $d=3$, the outcomes are closest to those on the circle graph, but on random regular networks the movement cost threshold that separates selection favoring cooperators from favoring defectors is higher: $\lambda = 0.65$ for the random regular networks vs.\ $\lambda = 0.45$ for the circle graph. As the degree of the networks increases, the region where selection favors cooperators expands, while the region where selection favors defectors shrinks; regions of neutral selection appear in between. The outcomes stabilize starting from $d=10$ and match those for the complete graph.

The fixation probabilities of cooperators remain stable for small movement costs regardless of the degree, but they increase with the average degree for larger movement costs (figures~\ref{fig:RRfast}(b) and \ref{fig:RRfast}(c)).

\subsection{Watts--Strogatz networks}

\begin{figure}
\centering
\includegraphics[width=\textwidth]{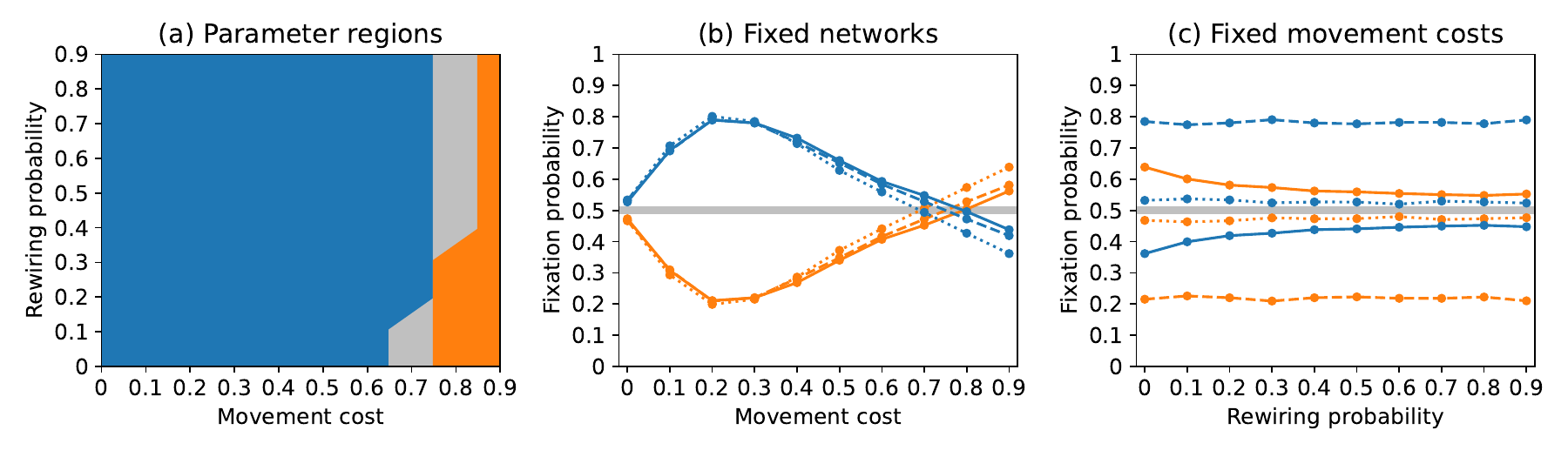} 
\caption{The outcomes in the non-rare interactive mutations scenario for Watts--Strogatz networks with fixed average degree $d=6$. (a) Parameter regions showing the possible outcomes. The results stabilize starting from $p=0.4$ but never match those of the complete graph.  (b) Equilibrium fixation probabilities of cooperators (blue) and defectors (orange) as functions of the movement cost. Each plot corresponds to a fixed value of the network parameter: $p=0$ for dotted lines, $p=0.2$ for dashed lines, $p=0.4$ for solid lines. (c) Equilibrium fixation probabilities of cooperators (blue) and defectors (orange) as functions of the network parameter. Each plot corresponds to a fixed value of the movement cost: $\lambda = 0$ for dotted lines, $\lambda = 0.3$ for dashed lines, $\lambda = 0.9$ for solid lines.}
\label{fig:WSd6fast}
\end{figure}

Figure~\ref{fig:WSd6fast} shows the outcomes for Watts--Strogatz networks with fixed initial degree $d=6$ of the nodes in the starting ring lattice and varying rewiring probability $p$. For small values of $p$, selection favors cooperators for small and intermediate movement costs, and selection is either neutral or favors defectors for larger movement costs. As the rewiring probability increases, the region where selection favors cooperators expands slightly, and the region where selection favors cooperators shrinks slightly. The outcomes stabilize at $p=0.4$, and they are similar to those for the complete graph. The only difference is that for Watts--Strogatz networks, selection is neutral for $\lambda = 0.8$.

The fixation probabilities remain relatively stable with respect to rewiring probability with fixation probabilities of cooperators increasing slightly with higher rewiring probabilities for larger movement costs (see figures~\ref{fig:WSd6fast}(b) and \ref{fig:WSd6fast}(c)).

\begin{figure}
\centering
\includegraphics[width=\textwidth]{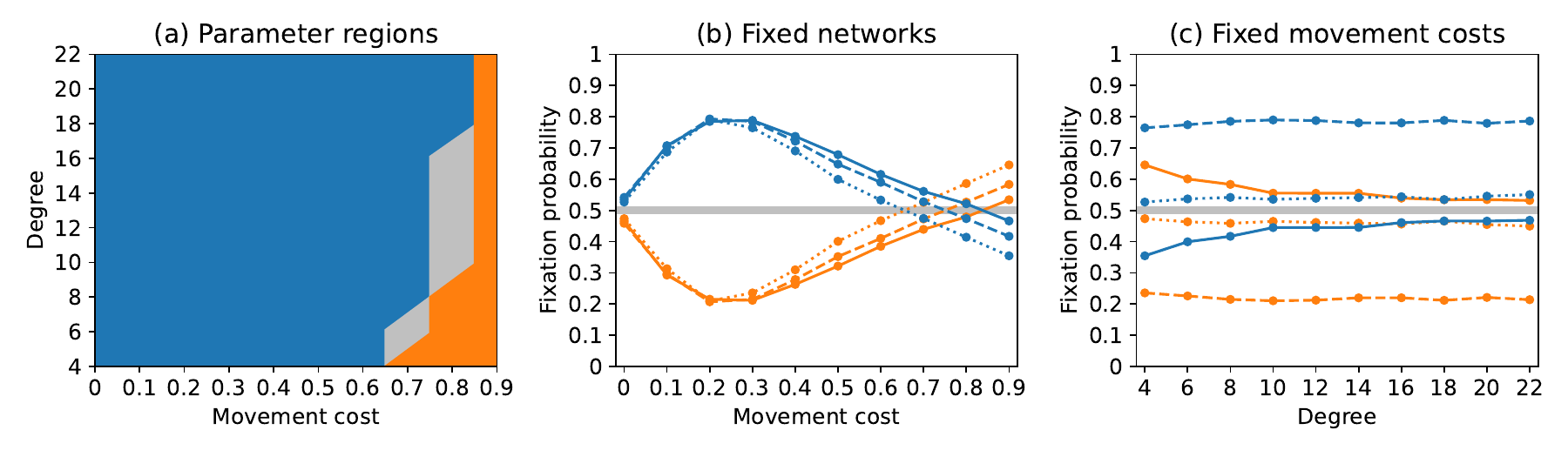} 
\caption{The outcomes in the non-rare interactive mutations scenario for Watts--Strogatz networks with fixed rewiring probability $p=0.1$. (a) Parameter regions showing the possible outcomes. The results match those of the complete graph starting from $d=18$.  (b) Equilibrium fixation probabilities of cooperators (blue) and defectors (orange) as functions of the movement cost. Each plot corresponds to a fixed value of the network parameter: $d=4$ for dotted lines, $d=8$ for dashed lines, $d=18$ for solid lines. (c) Equilibrium fixation probabilities of cooperators (blue) and defectors (orange) as functions of the network parameter. Each plot corresponds to a fixed value of the movement cost: $\lambda = 0$ for dotted lines, $\lambda = 0.3$ for dashed lines, $\lambda = 0.9$ for solid lines.}
\label{fig:WSp10fast}
\end{figure}

Figure~\ref{fig:WSp10fast} shows the outcomes for Watts--Strogatz networks with fixed rewiring probability $p=0.1$ and varying initial degree of nodes $d$ in the starting ring lattice. For networks with small average degrees, the outcomes are similar to those for the circle graph with selection favoring cooperators for smaller movement costs and defectors for larger movement costs. As the degree increases, the region where selection favors cooperators expands, and the region where selection favors defectors shrinks. The behavior stabilizes at $d=18$ where it becomes identical to that on the complete graph.

Increasing the average degree results in higher fixation probabilities for cooperators for larger movement costs (see figures~\ref{fig:WSp10fast}(b) and \ref{fig:WSp10fast}(c)).

\subsection{Summary}

Similarly to the rare interactive mutations scenario, in the non-rare interactive mutations scenario the outcomes resemble those for the complete graph as long as the average degree of the network is sufficiently high. But there are several differences from the previous scenario. First, the fixation probabilities were affected less by varying the network parameter in the non-rare interactive mutation case. Second, the qualitative outcomes didn't always completely match those of the complete graph (for example, on Barabási--Albert networks). Yet any qualitative differences were due to the fixation probabilities being close to the neutral one, and hence small differences in the actual fixation probabilities resulted in formal qualitative differences. Finally, the average degrees where the outcomes matched those of the complete graph were usually higher than those in the rare interactive mutations case.

\section{Discussion}\label{sec:discuss}

In this paper, we extended our investigation of the effect of network topology on the evolution of cooperation on evolving multiplayer networks \cite{Erovenko2019b}. We adapted the Markov movement model from complete, circle, and star graph to arbitrary networks. This allowed us to consider a wider range of network topologies. We concentrated on standard well-known types of random networks: Barabási--Albert, Erdős--Rényi, random regular, and Watts--Strogatz networks. We considered 10 different networks of each type, which were constructed by varying one network parameter. All networks had a fixed number of $50$ nodes to match the largest size graphs considered in \cite{Erovenko2019b}. We used the \texttt{networkx} package to create 10 sample networks for each value of the parameter; the simulations were performed on these sample networks.

We investigated the outcomes of the evolutionary process on these random networks in a population consisting of cooperators and defectors. Individuals explored the networks based on their exploration strategies (staying propensities) and the attractiveness of their current group, and played a public goods game with those they met. The fitness accumulated via these multiplayer games was used to evolve the population using the BDB process.

When the random networks had small average degrees, we saw a wide variation in the outcomes in the rare interactive mutations scenario similarly to \cite{Erovenko2019b}. The results for Barabási--Albert networks (of small average degree) were similar to those for the star graph; the results for Erdős--Rényi networks were similar to those for the complete graph; the results for the random regular networks were somewhat similar to those for the circle graph; and the results for the Watts--Strogatz networks were similar to those for the circle graph.

Yet when the average degree was sufficiently large, the results became identical to those for the complete graph. This is a natural outcome, but the actual value of the average degree when this occurred was much lower than that of the complete graph, and it ranged from $6$ to $12$ depending on the network. The complete graph can be characterized as a graph with the largest clustering coefficient and average degree and the smallest degree centralization and average shortest path length. The random networks that had identical outcomes to the complete graph tended to have small average path length and sufficiently high clustering coefficient or low degree centralization.

We have not observed the same variance of outcomes as in extreme topologies of the complete, circle, and star graphs in the non-rare interactive mutations case. Even in random networks of small average degree the outcomes never resembled those for the star graph. Additionally, even for networks of large average degree the outcomes didn't always completely match those of the complete graph. When they did match, it usually required a higher average degree than in the rare interactive mutations case.

While random networks provided us with a wider range of network topologies to investigate, we had little control over their topologies because they are constructed algorithmically. We plan to extend our investigation further by manually constructing networks that would gradually deviate from the extreme topologies of the complete and star graphs. For example, we may consider star-like networks that are constructed by joining several smaller stars or cliqued networks that are constructed by joining several smaller complete graphs. We are interested in discovering what minimal changes can be made to alter the outcome of the evolution of cooperation on such structures.

\bibliographystyle{plain}

\bibliography{multiplayer-networks}

\end{document}